\documentclass[preprint,aps]{revtex4}
\usepackage{mathrsfs}

\usepackage{graphicx}
\usepackage{multirow}
\usepackage{makecell}
\usepackage{booktabs}

\begin{document}

\title{Evolution of Incommensurate Superstructure and Electronic Structure with Pb Substitution in (Bi$_{2-x}$Pb$_{x}$)Sr$_2$CaCu$_2$O$_{8+\delta}$ Superconductors}

\author{Jing Liu$^{1,2}$, Lin Zhao$^{1,*}$, Qiang Gao$^{1,2}$, Ping Ai$^{1,2}$, Lu Zhang$^{1,2}$, Tao Xie$^{1,2}$, Jianwei Huang$^{1,2}$, Ying Ding$^{1,2}$, Cheng Hu$^{1,2}$, Hongtao Yan$^{1,2}$,  Chunyao Song$^{1,2}$, Yu Xu$^{1,2}$, Cong Li$^{1,2}$, Yongqing Cai$^{1,2}$, Hongtao Rong$^{1,2}$, Dingsong Wu$^{1,2}$, Guodong Liu$^{1}$, Qingyan Wang$^{1}$, Yuan Huang$^{1}$, Fengfeng Zhang$^{3}$, Feng Yang$^{3}$, Qinjun Peng$^{3}$, Shiliang Li$^{1,2,5}$, Huaixin Yang$^{1,2}$, Jianqi Li$^{1,2,5}$, Zuyan Xu$^{3}$ and Xingjiang Zhou$^{1,2,4,5,*}$}

\affiliation{
\\$^{1}$Beijing National Laboratory for Condensed Matter Physics, Institute of Physics, Chinese Academy of Sciences, Beijing 100190, China.
\\$^{2}$University of Chinese Academy of Sciences, Beijing 100049, China.
\\$^{3}$Technical Institute of Physics and Chemistry, Chinese Academy of Sciences, Beijing 100190, China.
\\$^{4}$Songshan Lake Materials Laboratory, Dongguan, Guangdong 523808, China.
\\$^{5}$Collaborative Innovation Center of Quantum Matter, Beijing 100871, China.
\\$^{*}$Corresponding author: LZhao@iphy.ac.cn,  XJZhou@iphy.ac.cn.
}

\date{\today}

\begin{abstract}
High-quality Bi$_{2-x}$Pb$_{x}$Sr$_2$CaCu$_2$O$_{8+\delta}$ (Bi2212) single crystals have been successfully grown by the traveling solvent floating zone technique with a wide range of Pb substitution ($x=0$--0.8). The samples are characterized by transmission electron microscope (TEM)  and measured by high resolution laser-based angle-resolved photoemission spectroscopy (ARPES) with different photon energies. A systematic evolution of the electronic structure and superstructure with Pb substitution has been revealed for the first time. The superstructure shows a significant change with Pb substitution and the incommensurate modulation vector ($\textbf{\emph{Q}}$) decreases with increasing Pb substitution. In the meantime, the superstructure intensity from ARPES measurements also decreases dramatically with increasing Pb concentration. The superstructure in Bi2212 can be effectively suppressed by Pb substitution and it nearly disappears with a Pb substitution of $x=0.8$.  We also find that the superstructure bands in ARPES measurements depend sensitively on the photon energy of lasers used; they can become even stronger than the main band when using a laser photon energy of 10.897~eV.  These results provide important information on the origin of the incommensurate superstructure and its control and suppression in bismuth-based high temperature superconductors.
\end{abstract}

\maketitle

\newpage

\section{\label{sec:level1}Introduction}

High temperature cuprate superconductors have been extensively studied for more than thirty years due to its unusually high critical temperature ($T_{\rm c}$), unique normal state, and superconducting properties\cite{1,2,3,4,5}.  The origin of the anomalous normal state and the high temperature superconductivity mechanism have not reached a consensus yet. Angle-resolved photoemission spectroscopy (ARPES)\cite{3,6,7,8} has played a key role in studying the electronic structure of the cuprate superconductors,  including revealing the distinct {d}-wave superconducting gap symmetry\cite{9,10,11,12},  the existence of pseudogap\cite{11,13,14,15}, and many-body effects\cite{16,17,18,19,20,21,22,23,24,25,26}.  Bi$_{2}$Sr$_2$CaCu$_2$O$_{8+\delta}$ (Bi2212) system, owing to the availability of high quality single crystals and readiness in  cleaving to get a clean and smooth surface, has long been the main workhorse to get these significant results by ARPES, as well as for scanning tunneling microscope/spectroscopy (STM/STS) measurements\cite{27,28,29,30,31}.  However, it is well-known that bismuth-based cuprate superconductors have incommensurate modulations in their crystal structure along the $b^{*}$ direction\cite{32,33,34,35}. This modulation gives rise to superstructure bands of various orders that significantly complicate the measured ARPES results, particularly near the important antinodal region\cite{36,37,38,39}.   A related issue under debate is about the origin of the superstructure bands: whether they are intrinsic that come  from the CuO$_2$ planes directly or extrinsic that are formed when the photoelectrons from CuO$_2$ planes are diffracted by the BiO layers\cite{37,38}.  It has been found that Pb-substitution in Bi2212 can effectively suppress the incommensurate modulations\cite{40,41,42} and corresponding superstructure bands in ARPES\cite{43,44,45,46,47}.  However,  systematic study of the effect of Pb-substitution on the incommensurate modulation and electronic structure in Bi2212 is still lacking.

In this paper, we report the growth of a series of Pb-substituted Bi$_{2-x}$Pb$_{x}$Sr$_2$CaCu$_2$O$_{8+\delta}$ (Pb-Bi2212) single crystals and investigations on the evolution of the superstructure and electronic structure with the Pb-substitution.  High quality Bi$_{2-x}$Pb$_{x}$Sr$_2$CaCu$_2$O$_{8+\delta}$  single crystals with various Pb contents ($x=0$--0.8) are prepared by traveling solvent floating zone method. The crystal structure is characterized by transmission electron microscope (TEM) to directly determine the incommensurate modulation. The Pb-substitution effect on the electronic structure of Bi2212 is investigated by high resolution laser-based ARPES.  We find that the superstructure bands in ARPES measurements exhibit strong photoemission matrix element effect; its intensity depends sensitively on the laser photon energy and can become even stronger than that of the main band.  With increasing Pb-substitution in Bi2212,  the incommensurate modulation vector ($\textbf{\emph{Q}}$) of the superstructure decreases while the modulation strength also significantly weakens, giving rise to an overall suppression of the superstructure.  Our results provide important information on the origin of the superstructure bands in ARPES measurements and on the tuning and control of the superstructure in Bi2212 superconductors.

\section{\label{sec:level2}EXPERIMENT}
The Bi$_{2.1(2.2)-x}$Pb$_{x}$Sr$_{2(1.8)}$CaCu$_{2}$O$_{8+\delta}$ single crystals were grown by the traveling solvent floating zone method using an infrared radiation furnace equipped with four 300~W halogen lamps\cite{48}.   The nominal compositions and growth rate are summarized in Table~1. A fast growing rate (0.5~mm/h) was used in order to minimize the loss of Pb due to the high-volatility of PbO.  The growth atmosphere was air. The typical size of the plate-like single crystals is about 5~mm$\times$6~mm except of the $x=0.8$ samples which is about 2~mm$\times$2~mm, as shown in Fig.~\ref{Fig01_sample}{(a)}. The real composition of the grown single crystals was determined by induction-coupled plasma atomic emission spectroscopy (ICP-AES) analysis  and the results are listed in Table~1.  The measured composition of the as-grown single crystals basically follows the nominal composition but the Pb content is slightly less  due to the loss of Pb during the growth process.  Compared with the stoichiometric formula of Bi$_2$Sr$_2$CaCu$_{2}$O$_{8+\delta}$, for all the samples, Pb$^{2+}$ not only substitutes at the Bi$^{3+}$ site but also goes to the Sr$^{2+}$ site that is consistent with previous Raman scattering studies\cite{42}.  For convenience, we will denote our single crystal samples by using their nominal Pb concentrations. The crystal structure of the Pb-Bi2212 single crystals was characterized  by x-ray diffraction (XRD) with Cu $K\alpha$ radiation ($\lambda=1.5418$~\AA). The results for four Pb-Bi2212 single crystals with different Pb contents  are shown in Fig.~\ref{Fig01_sample}{(b)}. The temperature dependence of magnetization was measured using a Quantum Design MPMS XL-1 system with a low magnetic field of 1 Oe. Both field cooling (FC), zero field cooling (ZFC), and AC field  were applied in order to get complementary information of their magnetization properties. The magnetic measurement results are shown in Figs.~\ref{Fig02_MPMSandPPMS}{(a)}--\ref{Fig02_MPMSandPPMS}{(e)}. The temperature dependence of the in-plane resistivity $\rho_{ab}$ was measured by  using the standard four-probe method and the results are shown in Fig.~\ref{Fig02_MPMSandPPMS}{(f)}. Electron diffraction and scanning transmission electron microscope (STEM) measurements were carried on by using JEM ARM200F TEM and the results are shown in Fig.~\ref{Fig03_TEM}.

High resolution angle-resolved photoemission measurements were performed by using a lab-based ARPES system equipped with 6.994~eV and 10.897~eV vacuum-ultra-violet (VUV) laser light sources and an angle-resolved time-of-flight electron energy analyzer (ARToF) with the capability of simultaneous two-dimensional momentum space detection\cite{8,49}.    One advantage of the ARToF analyzer is that it has much weaker non-linearity effect so that the measured signal is intrinsic to the sample. The energy resolution was set at 1~meV. The angular resolution was $\sim$0.1$^{\circ}$, corresponding to the momentum resolution of 0.002~{\AA}$^{-1}$ and 0.004~{\AA}$^{-1}$ at the photon energy of 6.994~eV and 10.897~eV, respectively. All the samples were cleaved \emph{in situ} at low temperature of 25 K and measured in ultrahigh vacuum with a base pressure better than 3$\times$10$^{-11}$ mbar. The ARPES measurements were carried out with both 6.994~eV laser (Fig.~\ref{Fig04_7eV}) and 10.897~eV laser (Fig.~\ref{Fig05_11eV}). In both measurements (Figs.~\ref{Fig04_7eV} and \ref{Fig05_11eV}), the electric field vector of the incident laser was perpendicular to the nodal direction ($(0,0)$--$(\pi,\pi)$) of the measured Bi2212 sample. The Fermi level is referenced by measuring on clean polycrystalline gold that is electrically connected to the sample or reference to the nodal direction of Bi2212 superconductors  where the superconducting gap is known to be zero.

\section{\label{sec:level3}RESULTS and discussion}

\begin{figure}[tbp]
\begin{center}
\includegraphics [width=0.8\columnwidth,angle=0]{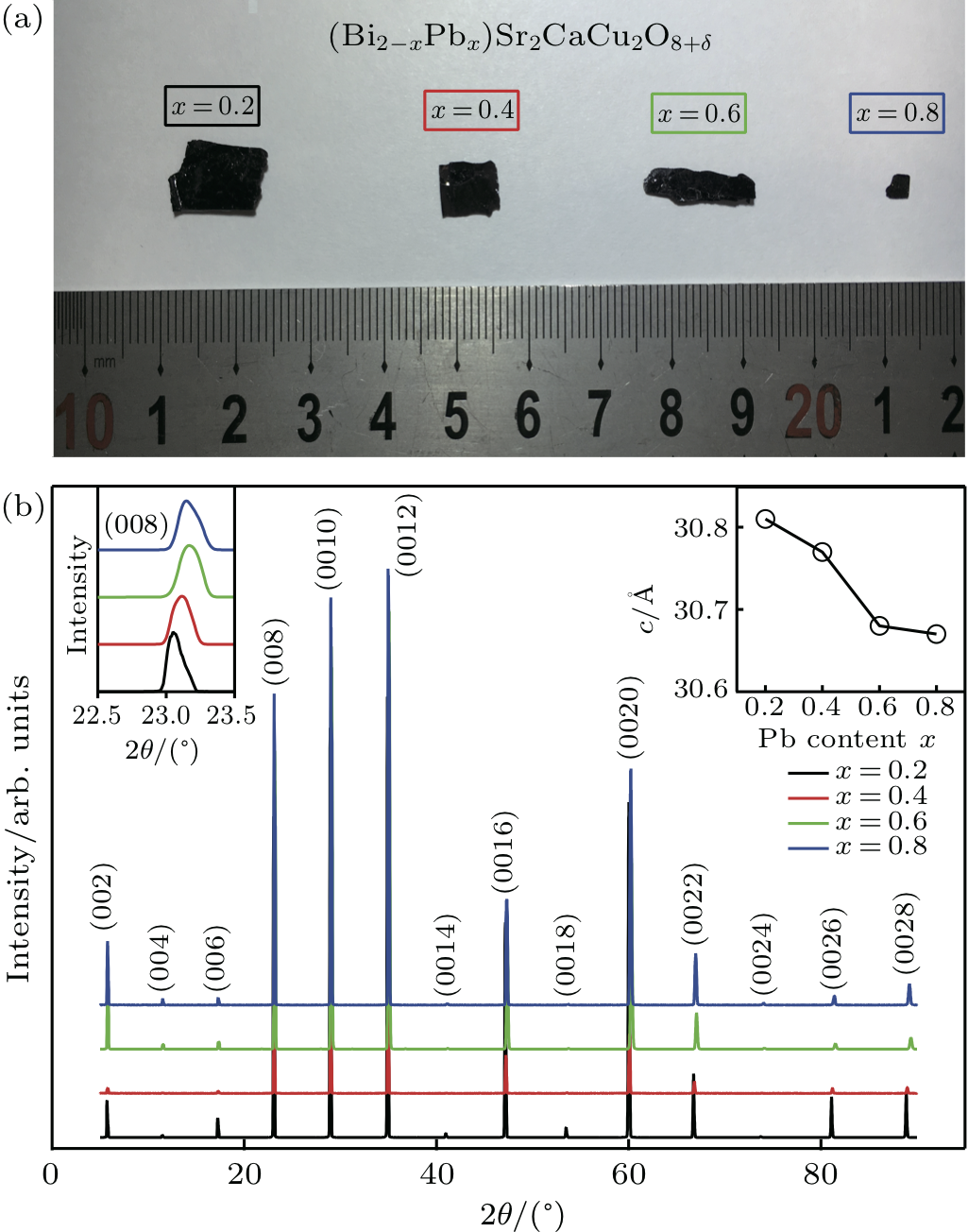}
\end{center}
\caption {\textbf{Pb-substituted Bi2212 single crystals and their structure characterization.} (a) Photos of Pb-Bi2212 single crystals cleaved from as-grown ingots with various nominal compositions. (b) XRD patterns for cleaved Pb-Bi2212 single crystals with different  Pb contents. The top-left inset shows the expanded (008) peak to highlight the peak width and its position variation with Pb content. The measured $c$-axis lattice constant is shown in the top-right inset.
}
\label{Fig01_sample}
\end{figure}

Figure~\ref{Fig01_sample}{(b)} shows the XRD patterns for the four  Pb-Bi2212 single crystals with different Pb contents. The measured surface is (001) $a$--$b$ plane and all the observed peaks can be indexed to the (00$l$) peaks of Bi2212, indicating a pure single-phase.  The peaks are sharp, as exemplified by the (008) peaks in the top-left inset of Fig.~\ref{Fig01_sample}{(b)} which have a width of $\sim$0.15$^\circ$ (full width at half maximum), indicating high crystallinity of the single crystals. The $c$-axis lattice constant is calculated according to these XRD patterns.  It exhibits a  monotonic decrease with the increasing Pb content, as shown in the top-right inset of Fig.~\ref{Fig01_sample}{(b)}.

\begin{figure*}[tbp]
\begin{center}
\includegraphics [width=1.0\columnwidth,angle=0]{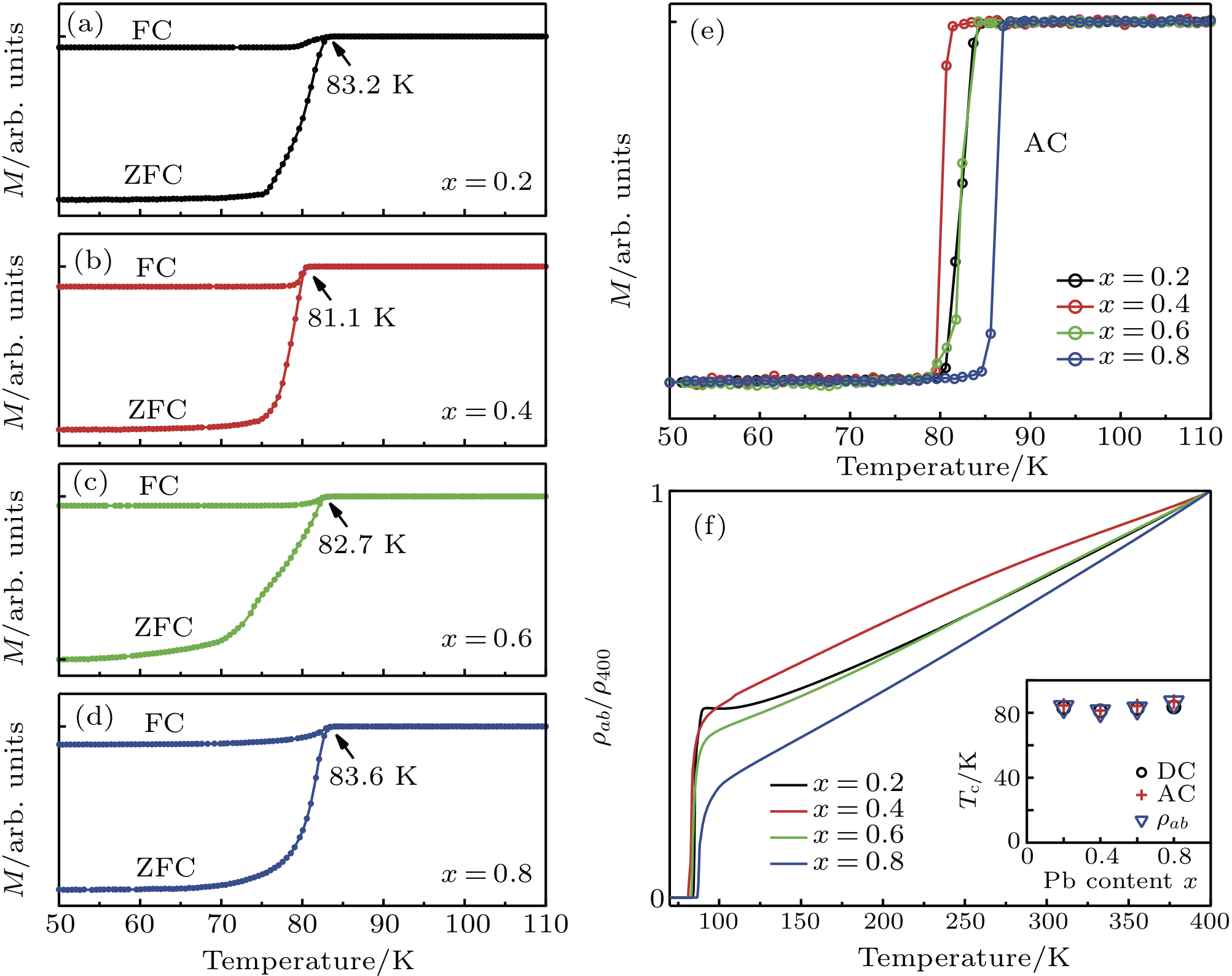}
\end{center}
\caption {\textbf{Transport and magnetic properties of Pb-Bi2212 single crystals with various Pb contents.} (a)--(d) Temperature dependence of DC magnetization measured under a magnetic field of 1 Oe for Pb-Bi2212 with different Pb contents. The onset superconducting transition temperature is marked in the figures.   (e) Temperature dependence of AC magnetization of the same samples  in (a)--(d). (f) Temperature dependence of the in-plane resistivity of Pb-Bi2212 single crystals with different Pb contents. The curves are normalized by the value at the temperature of 400~K.  The bottom-left inset shows the measured superconducting transition temperature ($T_{\rm c}$) of Pb-Bi2212 single crystals from (a)--(d) DC magnetization, (e) AC magnetization, and (f) resistivity measurements.
}
\label{Fig02_MPMSandPPMS}
\end{figure*}

 Figures~\ref{Fig02_MPMSandPPMS}{(a)}--\ref{Fig02_MPMSandPPMS}{(d)} show the temperature dependence of magnetization of Pb-Bi2212 single crystals with different Pb contents.  All the Pb-Bi2212 samples show clear superconducting transition in both FC and ZFC measurement modes. The measured superconducting transition temperatures (onset) are marked for each of the samples in Figs.~\ref{Fig02_MPMSandPPMS}{(a)}--\ref{Fig02_MPMSandPPMS}{(d)} and plotted in the inset of Fig.~\ref{Fig02_MPMSandPPMS}{(f)}.  For the same four samples,  we also measured their AC magnetic susceptibility, as shown in Fig.~\ref{Fig02_MPMSandPPMS}{(e)}.  The transition temperature measured by AC susceptibility (also plotted in the inset of Fig.~\ref{Fig02_MPMSandPPMS}{(f)}) is consistent with that measured  by FC and ZFC methods but with much narrower transition width because the AC measurement is less sensitive to the residual magnetic field in the magnetic measurement system.  The sharp superconducting transition (transition width of 1.5 K in Fig.~\ref{Fig02_MPMSandPPMS}{(e)}) indicates high quality of all the Pb-Bi2212 single crystals. The normalized in-plane resistivity $\rho_{ab}$ of these as-grown Pb-Bi2212 single crystals with various Pb contents is shown in Fig.~\ref{Fig02_MPMSandPPMS}{(f)}.  They all exhibit metallic behaviors and the samples appear to get  more metallic with increasing Pb content.  The superconducting transition temperature obtained from the resistivity measurements (zero resistance temperature) is plotted in the inset of Fig.~\ref{Fig02_MPMSandPPMS}{(f)}.  It is clear that for all the four  Pb-Bi2212 samples with different Pb contents, the measured $T_{\rm c}$s from DC magnetization (Figs.~\ref{Fig02_MPMSandPPMS}{(a)}--\ref{Fig02_MPMSandPPMS}{(d)}), AC magnetization (Fig.~\ref{Fig02_MPMSandPPMS}{(e)}), and resistivity measurements are in good agreement. It is also interesting to note that, although the actual Pb-substitution in these Pb-Bi2212 samples varies in a large range from 0.16 to 0.58 (Table~1), the superconducting transition temperature varies only in a very narrow range between 81 K and 84 K.   Since the as-grown pristine Bi$_2$Sr$_2$CaCu$_{2}$O$_{8+\delta}$ is usually close to be optimally-doped\cite{48},   Pb$^{2+}$ substitution into the Bi$^{3+}$ site tends to introduce extra holes into the Bi2212 samples, these Pb-Bi2212 samples should be therefore in the overdoped region, which is also supported by our ARPES results (Fig.~\ref{Fig05_11eV}).   In addition to the Pb-substitution effect, the doping level of the Bi2212 samples also depends on the oxygen content. Therefore, the nearly constant $T_{\rm c}$ for the Pb-Bi2212 samples with different Pb contents may be attributed to similar hole doping level that results from the balance between Pb-substitution and oxygen content during the growth process.

\begin{figure*}[tbp]
\begin{center}
\includegraphics [width=1.0\columnwidth,angle=0]{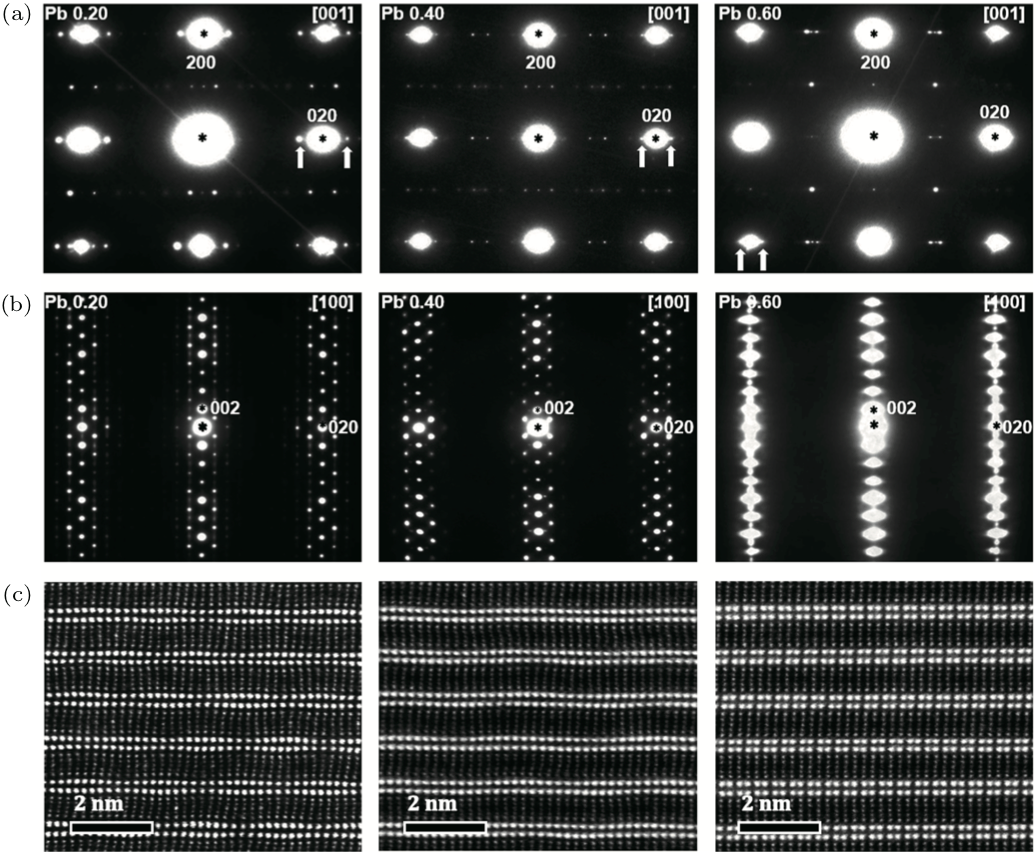}
\end{center}
\caption {\textbf{Electron diffraction patterns and STEM images of Bi2212 single crystals with various Pb contents.} (a) Selected-area diffraction patterns along $[001]$ zone-axis for $x=0.2$ (left panel), $x=0.4$ (middle panel), and $x=0.6$ (right panel) samples. (b) Selected-area diffraction patterns along $[100]$ zone-axis for the same samples. (c) $[100]$ zone-axis STEM images for the same samples.
}
\label{Fig03_TEM}
\end{figure*}

In order to investigate the crystal structure and the incommensurate modulations, systematic TEM studies were carried out on the series of Pb-Bi2212 samples. Figure~\ref{Fig03_TEM}{(a)} shows the electron diffraction results along the [001] zone-axis on  Pb-Bi2212 samples with three typical Pb concentrations $x=0.2$, 0.4, and 0.6.  The diffraction patterns show clear satellite spots around the main diffraction spots, as marked by two arrows in $x=0.2$ and $x=0.4$ samples around the main (020) spot, which are formed due to the formation of superstructure along the $b^\ast$ direction.  The satellite spot intensity relative to the main diffraction spots intensity and the distance of the two satellite spots along the $b^\ast$ direction are directly related to the modulation strength and the modulation period of the superstructure, respectively.  With the Pb content increasing from $x=0.2$ to $x=0.4$,  the satellite spot intensity decreases and becomes nearly unresolvable in the $x=0.6$ sample.  The distance between the two satellite spots decreases with the Pb content increasing from $x=0.2$ to $x=0.4$.  Similar results are observed in the diffraction patterns along the [100] zone-axis, as shown in Fig.~\ref{Fig03_TEM}{(b)}.  These results indicate that the superstructure in Bi2212 gets suppressed with Pb substitution by increasing the modulation period and decreasing the modulation strength.  These can be more directly seen from the [100] zone axis STEM images in Fig.~\ref{Fig03_TEM}{(c)}  where the BiO layers (bright horizontal double stripes), SrO layers,  CuO$_2$ layers, and Ca layers can all be clearly resolved.  Clear modulation can be seen in the BiO layers along the horizontal $b^\ast$ direction in the $x=0.2$ sample (left panel of Fig.~\ref{Fig03_TEM}{(c)}).  Such a modulation gets weaker in the $x=0.4$ sample  (middle panel of Fig.~\ref{Fig03_TEM}{(c)}), and nearly invisible in the $x=0.6$ sample  (right panel of Fig.~\ref{Fig03_TEM}{(c)}),  consistent with the superstructure suppression with Pb substitution observed in electron diffractions (Figs.~\ref{Fig03_TEM}{(a)} and \ref{Fig03_TEM}{(b)}).  We also note by a close inspection of Fig.~\ref{Fig03_TEM}{(c)} that  the incommensurate modulation exists not only in the BiO layers, but also in the SrO  and  CuO$_2$ layers\cite{32,50}.

The existence of the incommensurate modulations in Bi2212 gives rise to superstructure bands in the measured electronic structure, i.e.,  extra replica bands will be formed by shifting the original Fermi surface by $\pm n \textbf{\emph{Q}}$, where $\textbf{\emph{Q}}$ is the vector of the incommensurate modulation and $n$ is the order of the superstructure bands\cite{36,37,38,39}.   Such an effect also renders ARPES as an alternative technique to detect the superstructure in Bi2212.  To systematically investigate the evolution of superstructure with Pb substitution in Pb-Bi2212, we performed high-resolution ARPES measurements on a series of Pb-Bi2212 samples with different Pb contents.

\begin{figure*}[tbp]
\begin{center}
\includegraphics [width=1.0\columnwidth,angle=0]{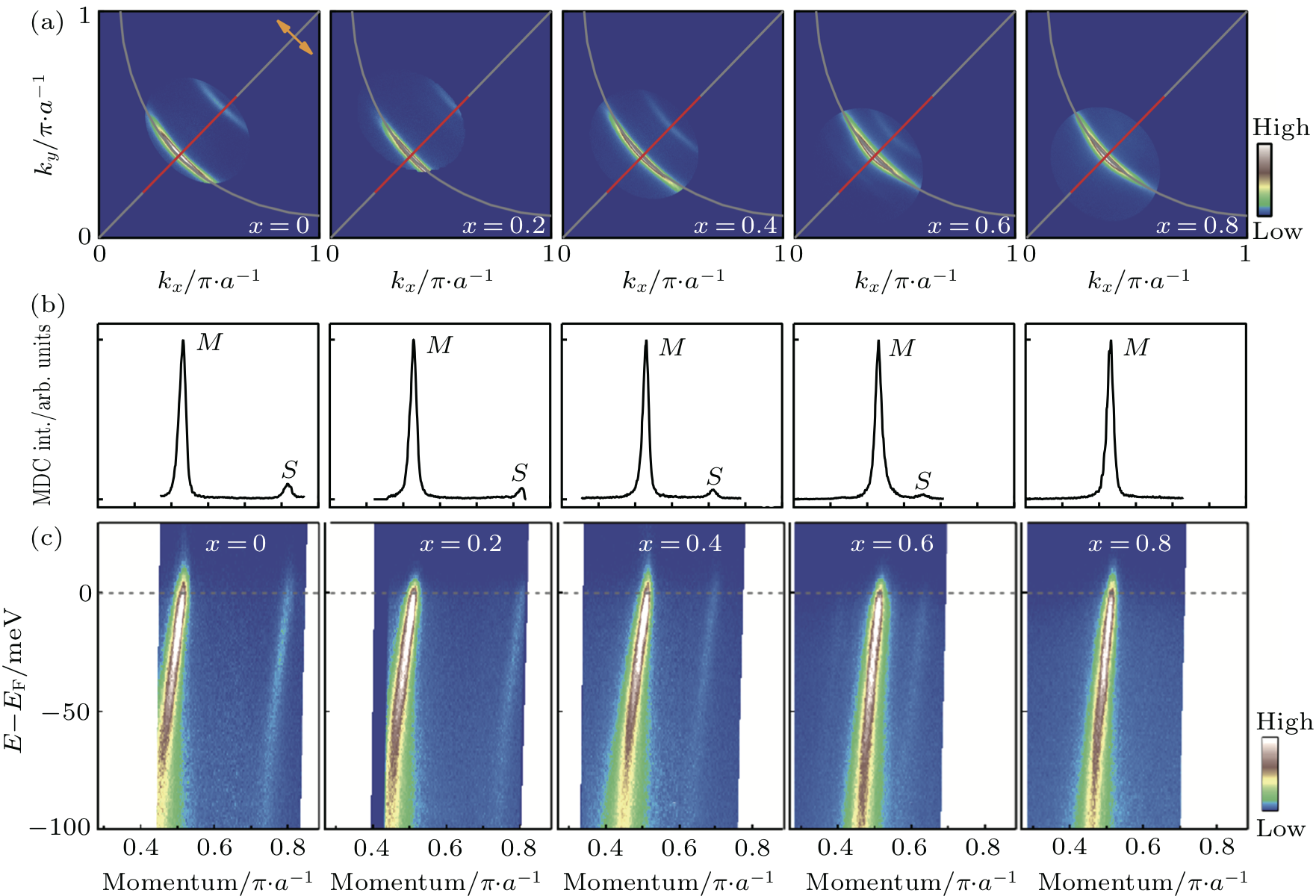}
\end{center}
\caption {\textbf{Fermi surface mapping and band structure of Pb-Bi2212 measured with 6.994 eV laser. } (a) Fermi Surface mappings for Pb-Bi2212 with various Pb contents (left to right panels correspond to $x=0$,  0.2, 0.4, 0.6, and 0.8,  respectively) measured  at a temperature of  25 K using a laser of photon energy 6.994 eV. Here $x=0$ sample is optimally doped Bi2212 with   $T_{\rm c}=91$ K. Each image is obtained by integrating measured spectral weight within $[-1, 1]$ meV energy window with respect to the Fermi level as a function of $k_x$ and $k_y$. The grey solid lines are the guides to the main Fermi surface. Orange double arrows represent the electric field vector direction of the incident laser. (c) Band structures measured along the nodal direction for the five  samples. The locations of the momentum cuts are marked in (a) by red lines. The corresponding momentum distribution curves at the Fermi level are shown in (b).  Two peaks are observed where $M$ peak represents the main band while $S$ peak represents the superstructure band.
}
\label{Fig04_7eV}
\end{figure*}

\begin{figure*}[tbp]
\begin{center}
\includegraphics [width=1.0\columnwidth,angle=0]{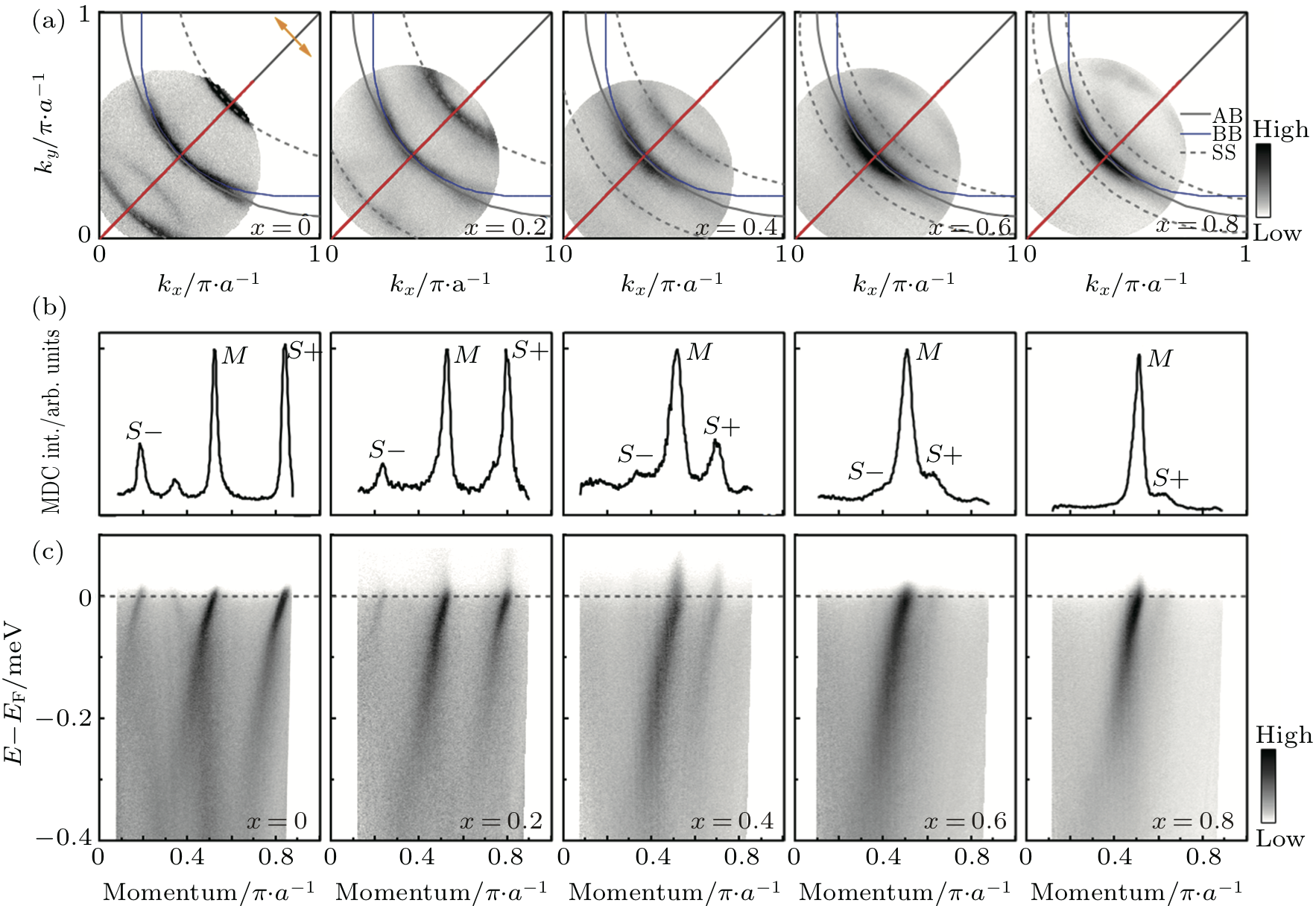}
\end{center}
\caption {\textbf{Fermi surface mapping and band structure of Pb-Bi2212 measured with 10.897 eV laser. } (a) Fermi surface mappings for Pb-Bi2212 with various Pb contents (left to right panels correspond to $x=0$,  0.2, 0.4, 0.6, and 0.8,  respectively) measured  at a temperature of  25 K using a laser of photon energy 10.897 eV. Here $x=0$ sample is overdoped Bi2212 with $T_{\rm c}=81$~K. Each image is obtained by integrating measured spectral weight within $[-1, 1]$ meV energy window with respect to the Fermi level as a function of $k_x$ and $k_y$. Orange double arrows represent the electric field vector direction of the incident laser. The grey and blue solid lines are the guides to the main antibonding (AB) and bonding (BB) Fermi surface sheets, respectively, while the dashed lines represent the first-order superstructure (SS) replicas. (c) Band structures measured along the nodal direction for the five   samples. The locations of the momentum cuts are marked in (a) by red lines. The corresponding momentum distribution curves at the Fermi level are shown in (b).  The $M$ peak represents the main band while $S-$ and $S+$  peaks represent two superstructure bands on the left and right sides of the main band, respectively.
}
\label{Fig05_11eV}
\end{figure*}

\begin{figure*}[tbp]
\begin{center}
\includegraphics [width=1.0\columnwidth,angle=0]{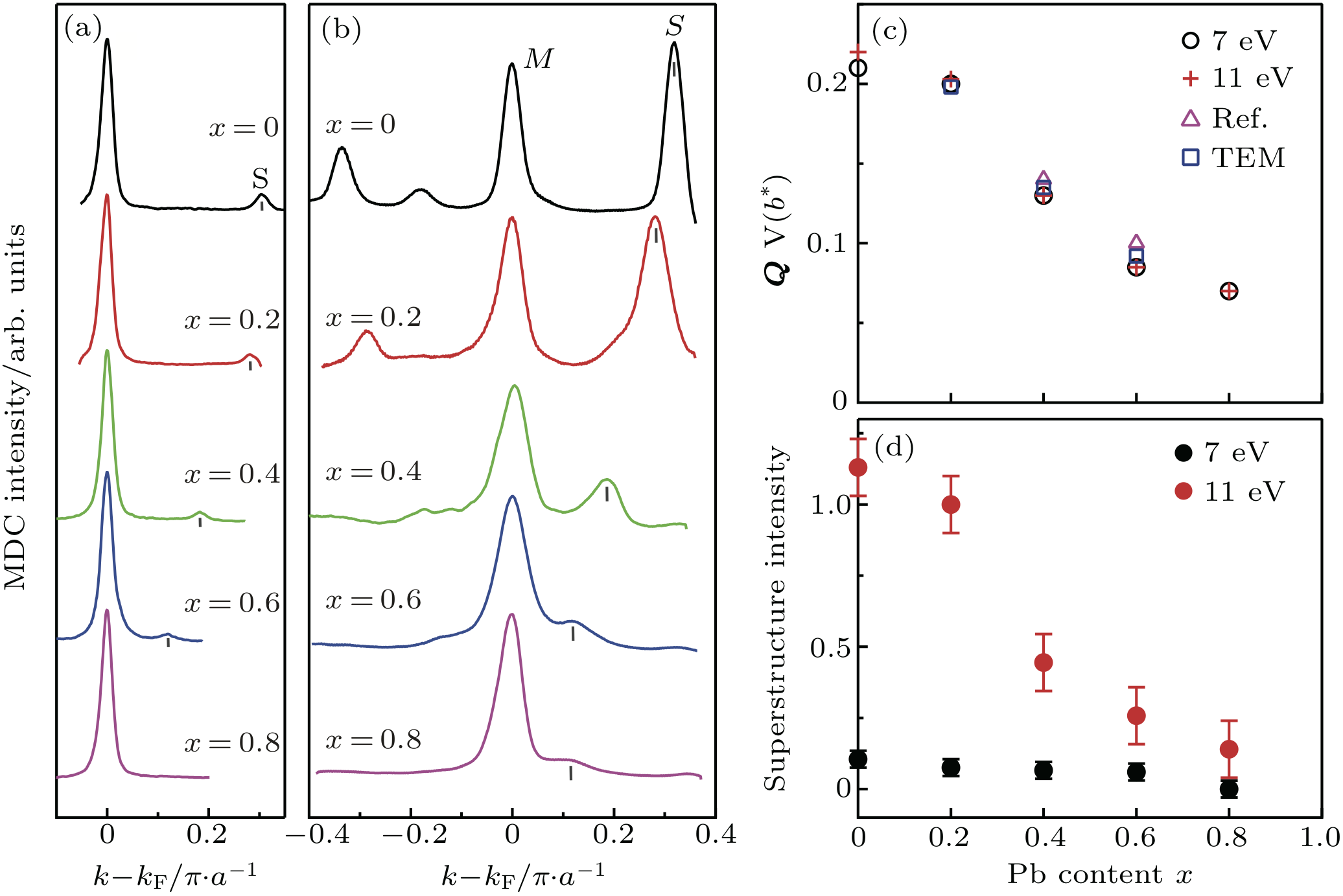}
\end{center}
\caption {\textbf{Evolution of superstructure bands with Pb substitution in Pb-Bi2212.}  Panles (a) and (b) compare MDCs along the nodal direction at the Fermi level for Pb-Bi2212 samples with different Pb contents measured by (a) 6.994 eV laser (like in Fig.~\ref{Fig04_7eV}) and (b)  10.897 eV laser (like in Fig.~\ref{Fig05_11eV}).   The peaks of superstructure bands are marked with black short lines. (c) Variation of the determined $\textbf{\emph{Q}}$ vector with Pb content in Pb-Bi2212. Black circle and red cross are obtained from laser-ARPES measurements with 6.994 eV laser and 10.897 eV laser, respectively. Blue  rectangular is obtained from TEM measurements (Fig.~\ref{Fig03_TEM}). Pink triangle is  from TEM measurement in Ref.~\cite{40}. (d)  Variation of superstructure band intensity with Pb content in Pb-Bi2212 measured by laser-ARPES with 6.994 eV laser (black dots) and 10.897 eV laser (red dots). The intensity is normalized to that of the main band.
}
\label{Fig06_SuperstructureHQ}
\end{figure*}

Figure~\ref{Fig04_7eV} shows the Fermi surface and band structure of Pb-Bi2212 samples with five different Pb contents ranging from $x=0$ to $x=0.8$ measured at 25~K using 6.994~eV laser light source.  With the capability to simultaneously cover the two-dimensional momentum space of our ARToF-based laser  ARPES system, we can cover the main Fermi surface and the superstructure replica under the same experimental condition and obtain the measured data aligned precisely near the nodal region. This makes it possible to measure accurately the incommensurate modulation vector  ($\textbf{\emph{Q}}$) that is the momentum shift between the main Fermi surface and the first-order replica Fermi surface along the $b^\ast$ direction ((0,0)--($\pi$, $\pi$) diagonal direction in Fig.~\ref{Fig04_7eV}{(a)}). The strength of the structural modulation can also be measured by the intensity of the superstructure band relative to that of the main band. Clear superstructure Fermi surface is observed on the right side of the main Fermi surface  in the $x=0$ sample, which represents the first-order ($n=1$) Fermi surface replica caused by the incommensurate modulation (left panel in Fig.~\ref{Fig04_7eV}{(a)}). With increasing Pb-substitution, the distance between the main Fermi surface and the superstructure Fermi surface gets smaller, accompanied by the intensity decrease of the superstructure Fermi surface. The superstructure Fermi surface becomes invisible in the $x=0.8$ sample (right panel in Fig.~\ref{Fig04_7eV}{(a)}). These results indicate the gradual suppression of the superstructure modulation with   Pb-substitution which agrees well with the above TEM and STEM measurements (Fig.~\ref{Fig03_TEM}).   In order to quantitatively determine the characteristics of the superstructure bands, we extracted the band structure along the nodal direction (locations of the momentum cuts are marked by red lines in Fig.~\ref{Fig04_7eV}{(a)}), as  shown in Fig.~\ref{Fig04_7eV}{(c)} for these five Pb-Bi2212 samples. The corresponding momentum distribution curves (MDCs) at the Fermi level are shown in Fig.~\ref{Fig04_7eV}{(b)}. The magnitude of the incommensurate modulation vector $\textbf{\emph{Q}}$, determined from the distance between the M peak and S peak in each MDC, is plotted in Fig.~\ref{Fig06_SuperstructureHQ}{(c)}. The intensity of the superstructure band, determined by the intensity of the $S$ peak area relative to  the $M$ peak area, is plotted in Fig.~\ref{Fig06_SuperstructureHQ}{(d)}.

Figure~\ref{Fig05_11eV} shows Fermi surface mapping and band structure measurement on the Pb-Bi2212 samples with 10.897~eV laser photon energy. The larger photon energy makes it possible to simultaneously cover large momentum  space, in this case, the superstructure Fermi surface sheets on both the left  and right sides of the main Fermi surface are covered. Also both the antibonding band (grey solid line) and bonding band (blue solid line) are resolved due to the bilayer splitting in Bi2212\cite{45,51}.  This observation further supports that all the Pb-Bi2212 samples are in the overdoped region because we did not observe bilayer splitting in optimally and underdoped Bi2212 under the same measurement condition. For comparison, we also measured pristine overdoped Bi2212 sample ($x=0$)  with  $T_{\rm c}=81$~K that is obtained by annealing and has a similar $T_{\rm c}$ as those of Pb-Bi2212 samples. Figure~\ref{Fig05_11eV}{(c)} shows the band structure along the nodal direction (the locations of the momentum cuts are shown in Fig.~\ref{Fig05_11eV}{(a)} by red lines) and the corresponding MDCs at the Fermi level are shown in Fig.~\ref{Fig05_11eV}{(b)}.  The superstructure band in the 10.897~eV laser ARPES measurement is significantly enhanced in intensity when compared with the 6.994~eV laser ARPES measurement (Fig.~\ref{Fig04_7eV}), which is due to the photoemission matrix element effect\cite{3}.  The superstructure bands are comparable in intensity to the main band in the $x=0.2$ sample, and can become even stronger than the main band in the $x=0$ sample (Fig.~\ref{Fig05_11eV}{(b)}).  This giant signal enhancement  also makes it possible to detect the superstructure band even when it becomes rather weak.  As seen in Fig.~\ref{Fig05_11eV},  the evolution of the superstructure band with the Pb substitution shows an overall agreement with that measured by 6.994~eV laser ARPES (Fig.~\ref{Fig04_7eV}), i.e.,  the modulation vector gets smaller and the intensity of the superstructure band gets weaker with increasing Pb substitution in Bi2212.  However, in our present 10.897~eV ARPES measurement, the superstructure band is still visible in the $x=0.8$ sample, which is not resolvable in the 6.994~eV laser ARPES measurement. In the TEM measurements (Fig.~\ref{Fig03_TEM}), the superstructure becomes hard to resolve even in the $x=0.6$ sample. These results demonstrate that ARPES  has provided an alternative, but more sensitive and quantitative measurement on the superstructure in Bi2212.  Our 10.897~eV laser ARPES measurements  indicate that the incommensurate modulation still survives in the $x=0.8$ sample although it has become significantly weakened and not resolvable in the TEM and 6.994~eV laser ARPES measurements.

Figure~\ref{Fig06_SuperstructureHQ} summarizes the main results on the superstructure evolution with Pb substitution in Pb-Bi2212 samples from the TEM and ARPES measurements. For a direct comparison, figures~\ref{Fig06_SuperstructureHQ}{(a)} and   \ref{Fig06_SuperstructureHQ}{(b)} show MDCs at the Fermi level along the nodal direction for the Pb-Bi2212 samples with different Pb contents measured using 6.994~eV and 10.897~eV laser photon energies, respectively.  The position and intensity variation of the superstructure bands with the Pb substitution relative to the main band can be clearly seen. In particular, the intensity enhancement of the superstructure band in the 10.897~eV ARPES measurement becomes apparent compared to the 6.994~eV ARPES measurement. Figure~\ref{Fig06_SuperstructureHQ}{(c)} compiles the incommensurate modulation vector measured on Pb-Bi2212 samples by our TEM and ARPES measurements and previous TEM measurement\cite{40} which give consistent results.  The modulation vector shows a monotonic decrease from (0.21, 0.21) for the $x=0$ pristine Bi2212 to (0.07, 0.07) for the $x=0.8$ sample.  The superstructure band intensity,  denoted by the area ratio between the superstructure peak and the main peak in MDCs (Figs.~\ref{Fig06_SuperstructureHQ}{(a)} and \ref{Fig06_SuperstructureHQ}{(b)}),  is shown in Fig.~\ref{Fig06_SuperstructureHQ}{(d)} for 6.994~eV and 10.897~eV ARPES measurements. It also falls monotonically with the increasing Pb substitution content although the intensity in the 10.897~eV measurement is significantly stronger than that in the 6.994~eV measurement.

It is under debate on the origin of the superstructure bands in Bi2212 in ARPES measurements, whether it is intrinsic that comes directly from the CuO$_2$ planes or extrinsic that is formed when the photoelectrons from CuO$_2$ planes get diffracted from the superstructure in the BiO layers\cite{37,38}.  Our present results provide more information on this issue and favors the former intrinsic scenario. If the superstructure band is produced by the diffraction of photoelectrons from CuO$_2$ planes through the incommensurate modulation in BiO layers, the mechanism is similar to that of transmission electron microscope. In this case, as seen in the TEM diffraction patterns in Fig.~\ref{Fig03_TEM}, the diffracted signal is only a small fraction of the undiffracted transmitted signal. In terms of the diffraction model, one may think the photoelectrons from CuO$_2$ planes as the TEM electron source, and the BiO layers as the TEM sample. It is difficult to understand how the superstructure band can become even stronger than the main band, as we observed in the 10.897~eV ARPES measurement on the $x=0$ Bi2212 sample (Fig.~\ref{Fig05_11eV}{(a)}). Furthermore, the relative intensity between the superstructure band and the main band should be insensitive to the photoelectron energy like the electron diffraction in TEM; this is obviously inconsistent with our results. We note that, since the photoelectrons from CuO$_2$ planes must pass through other layers to get out of the sample, the superstructure modulation in BiO layers or other layers will play some role in diffracting the photoelectrons. But this diffraction effect cannot be dominant in producing the superstructure bands in Bi2212 superconductors.

The signal of the superstructure bands is more likely generated from the CuO$_2$ planes themselves. From direct structural characterization (Fig.~\ref{Fig03_TEM} and Ref.~\cite{50}), there are modulations in the CuO$_2$ planes with a periodicity that is similar to that in BiO layers. The modulation in this case has become an integral part of the CuO$_2$ plane. First, the modulation will produce superstructure bands that are intrinsic to the CuO$_2$ plane. Second, the intensity of the superstructure bands, as well as the main bands, can be affected by the photoemission matrix element effect\cite{3}. This can explain the photon energy dependence of the superstructure band intensity relative to the main band intensity, as we have observed. Since the photoemission matrix element is related to many factors like photon energy, photon polarization, and the energy and momentum of photoelectrons, this can also explain the intensity difference of the $S+$ and $S-$ superstructure bands (Fig.~\ref{Fig05_11eV}). This is consistent with the results that the relative intensity change of the main band and the shadow band in Bi2212 is strongly affected by the photoemission matrix element effects\cite{52}.  Quantitative understanding of the results asks for further calculation of the photoemission matrix element effect by taking into account of the superstructure bands in Bi2212\cite{53}.

Our systematic study of Pb substitution effect in Bi2212 can also shed some light on the formation mechanism of the superstructure in bismuth-based cuprate superconductors.  Incommensurate modulation structure is common in bismuth-based cuprates due to lattice mismatch between different layers in the crystal structure. Several possible formation mechanisms have been proposed including doping effect, element substitution effect, and extra oxygen atoms in the BiO layers\cite{54,55}.   It has been found that, potassium deposition on the surface of Bi2212 can effectively vary the carrier doping level over a large range, but it has little effect on the superstructure band\cite{56},  thus ruling out the effect of doping effect on the superstructure formation in Bi2212.  It is also found that, vacuum annealing of Bi2212 can vary its oxygen content as well as the doping level over a wide range, but the superstructure modulation does not show obvious change in the process\cite{57}.  This can rule out the extra oxygens as the main factor in the formation of superstructure modulation in Bi2212.  La can substitute Sr in the SrO layers in Bi$_2$Sr$_{2-x}$La$_x$CuO$_{6+\delta}$ (La-Bi2201) over a wide range ($x=0$--1.1), but it has little effect on the superstructure\cite{58,59}.   This indicates that the SrO layers do  not play the dominant role in the superstructure formation in La-Bi2201.  Pb substitution has been proven to be the most effective way  so far in controlling the  superstructure in Bi2212 and Bi2201\cite{60,61,62}.  Our present results have provided detailed information on the evolution of the superstructure with Pb substitution and  proven the significant role in suppressing the superstructure by Pb substitution.  According to the composition analysis in Table~\ref{table},  Pb mainly substitutes into the Bi sites in the BiO layers although some of them can also goes into the Sr sites in the SrO layers. But the composition of Sr keeps nearly unchanged when the Pb content varies from 0.16 to 0.58. These results indicate that the Pb substitution in the BiO layers plays the dominant role in controlling the superstructure formation in the bismuth-based cuprate superconductors.

\section{\label{sec:level4}conclusion}

In summary, by  Pb substitution into the Bi$_{2}$Sr$_{2}$CaCu$_{2}$O$_{8+\delta}$ superconductors,  we have prepared a series of  Pb-Bi2212 single crystals with different Pb contents.  The evolution of the superstructure in Pb-Bi2212 with Pb-substitution has been systematically studied by TEM  and ARPES measurements. The superstructure gets suppressed with increasing Pb substitution, manifested by the decrease of the incommensurate modulation vector as well as the modulation strength. We also find that the superstructure band exhibits strong sensitivity to the laser photon energy used in the ARPES measurements. These results provide important information on the origin of the superstructure formation in ARPES measurements, and on the tuning and control of the superstructure in Bi2212 samples. These information will also facilitate future ARPES studies on bismuth-based superconductors by preparing ideal samples and selecting proper experimental conditions.

\vspace{3mm}

\section{\label{sec:level5}Acknowledgments}
This work is supported by the National Key Research and Development Program of China (Grant No. 2016YFA0300300 and 2017YFA0302900), the Strategic Priority Research Program (B) of the Chinese Academy of Sciences (Grant No. XDB07020300 and XDB25000000), the National Natural Science Foundation of China (Grant No. 11334010 and 11534007), and the Youth Innovation Promotion Association of CAS (Grant No. 2017013).

\vspace{3mm}

$^{*}$Corresponding author:  LZhao@iphy.ac.cn,  XJZhou@iphy.ac.cn.



\bibliographystyle{unsrt}

\renewcommand\figurename{Fig.}

\newpage

\begin{center}
{\footnotesize{\bf Table 1. }Nominal and measured compositions, growth condition, {\it c}-axis lattice constant and T$_c$(onset) of  Pb-Bi2212 single crystals.\\
\vspace{2mm}
\begin{tabular}{ccccc}
\hline
Nominal composition   & Measured composition    & Growth rate (mm/hour)    & {\it c} (\AA)    & T$_c$(onset) (K) \\
\hline
$Bi_{1.9}Pb_{0.2}Sr_2CaCu_2O_{8+\delta}$      & $Bi_{1.92}Pb_{0.16}Sr_{1.8}CaCu_2O_{8+\delta}$  &0.5    & 30.81    & 83 \\
$Bi_{1.8}Pb_{0.4}Sr_{1.8}CaCu_2O_{8+\delta}$    & $Bi_{1.8}Pb_{0.3}Sr_{1.88}CaCu_2O_{8+\delta}$  &0.5    & 30.78    & 81 \\
$Bi_{1.6}Pb_{0.6}Sr_{1.8}CaCu_2O_{8+\delta}$    & $Bi_{1.68}Pb_{0.45}Sr_{1.84}Ca_{1.04}Cu_2O_{8+\delta}$  &0.5    & 30.68    & 83 \\
$Bi_{1.4}Pb_{0.8}Sr_{1.8}CaCu_2O_{8+\delta}$    & $Bi_{1.7}Pb_{0.58}Sr_{1.86}CaCu_2O_{8+\delta}$  &0.5    & 30.67    & 84 \\
\hline
\label{table}
\end{tabular}}
\end{center}

\end{document}